\documentclass[epj,english]{svjour}

\usepackage{graphics}
\usepackage{amsmath}
\usepackage{amssymb}

\begin{document}

\title{Comparison of Raman spectra and vibrational density of states between graphene nanoribbons with different edges}

\author{Sami Malola\inst{1 \and}\thanks{sami.malola@phys.jyu.fi} \and Hannu H\"akkinen\inst{1,2} \and Pekka Koskinen\inst{1}}

\institute{NanoScience Center, Department of Physics, FIN-40014 University of Jyv\"askyl\"a, Finland \and NanoScience Center, Department of Chemistry FIN-40014 University of Jyv\"askyl\"a, Finland}
 
\date{Received: date / Revised version: date}

\abstract{
Vibrational properties of graphene nanoribbons are examined with density functional based tight-binding method and non-resonant bond polarization theory. We show that the recently discovered reconstructed zigzag edge can be identified from the emergence of high-energy vibrational mode due to strong triple bonds at the edges. This mode is visible also in the Raman spectrum. Total vibrational density of states of the reconstructed zigzag edge is observed to resemble the vibrational density of states of armchair, rather than zigzag, graphene nanoribbon. Edge-related vibrational states increase in energy which corroborates increased ridigity of the reconstructed zigzag edge.
\PACS{
{61.46.-w}{Structure of nanoscale materials} \and
{64.70.Nd}{Structural transitions in nanoscale materials} \and
{63.22-m}{Phonons or vibrational states in low-dimensional structures and nanoscale materials} \and
{78.30Na}{Infrared and Raman spectra, fullerenes and related materials}
}
}

\titlerunning{Comparison of Raman spectra and VDOS of graphene nanoribbons with different edges}
\authorrunning{ S. Malola et al. }

\maketitle

In past decades carbon nanomaterials have shown their rich properties in several applications\cite{Science_297_787,nnano_2_605}. Here graphene nanoribbons are not an exception, and their use in many applications, among which transistors in nanoelectronics, have been investigated considering different edge structures\cite{APL_89_203107,APL_91_073103}. Low dimensionality and edges make ribbons particularly fascinating both for theory and applications.

The precise edge structure in graphene nanoribbons affects many properties like chemical reactivity\cite{JCP_126_134701}, electronic structure\cite{PRB_54_17954} and vibrations\cite{PRB_77_054302}. Vibrational properties play a role in structural stability \cite{PSS_245_695}, structure identification\cite{SSC_143_47} and ballistic transport through electron-phonon coupling\cite{SSC_143_47}. In structure identification scanning tun\-ne\-ling microscopy\cite{nnano_149_1,PRB_73_125415} can reach near atom resolution but ana\-lysis of the full edge structure and properties is often ambiguous. In this case Raman spectroscopy\cite{APL_91_173108,PRL_93_047403,SSC_143_47} is a valuable tool.

Vibrational properties of graphene nanoribbons have been recently studied for acoustic and optical phonons, symmetries, and Raman activity as a function of ribbon width\cite{APL_91_173108,PSS_245_695}. In this paper we consider recently reported self-passivating edge reconstruction of zigzag ribbon \cite{reconstru}, and investigate how the reconstruction affects Raman spectra and vibrational density of states (VDOS). Edge-loca\-lized contributions to Raman spectra and vibrational density of states change. It turns out that high-energy modes due to triple-bond vibrations, as well as overall changes in the vibrational density of states and changes in rigidity of the edges, make the reconstruction identifiable and visible in Raman spectra.
\begin{figure}[ht!!]
\centering
\resizebox{0.3516\vsize}{!}{\includegraphics{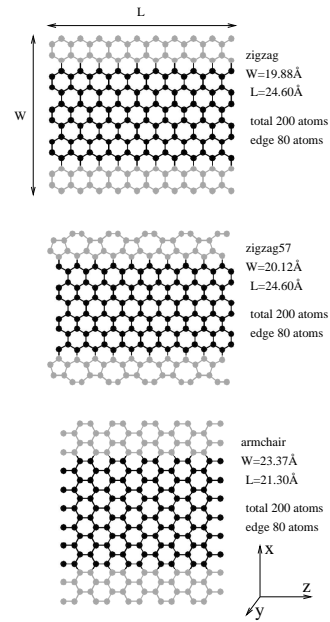}}
\caption{Examined structures of zigzag, reconstructed zigzag (zigzag57) and armchair graphene nanoribbons. Selected edge atoms are drawn in gray color. Coordinate axis is representated at lower right corner, where the ribbon direction (=periodic direction) is z-direction. $W=$width in non-periodic and $L=$ length in periodic direction.}
\label{fig1}
\end{figure}

\begin{figure*}[ht!!]
\centering
\resizebox{0.95\hsize}{!}{\includegraphics{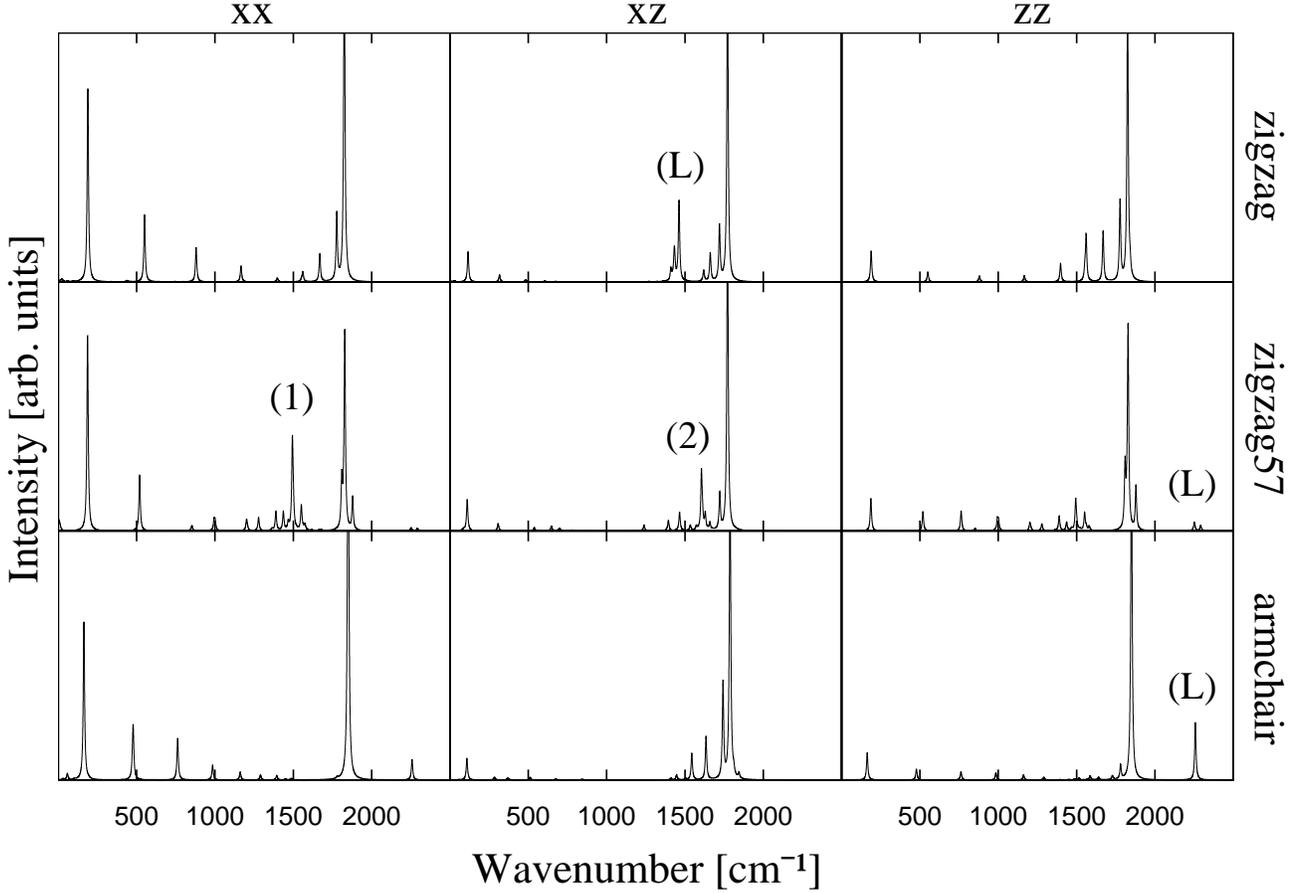}}
\caption{Raman spectra of zigzag, reconstructed zigzag (zigzag57) and armchair graphene nanoribbons (different rows). Columns denote different incident and scattered light polarization spectra (uppermost symbols xx, xz and zz). Edge-localized modes are assigned with symbol (L). Symbols (1) and (2) are refered to in the text. Raman spectra are drawn using Lorenz distribution with full width at half maximum of $5$~cm$^{-1}$.}
\label{fig2}
\end{figure*}

\begin{figure}
\centering
\resizebox{0.55\vsize}{!}{\includegraphics{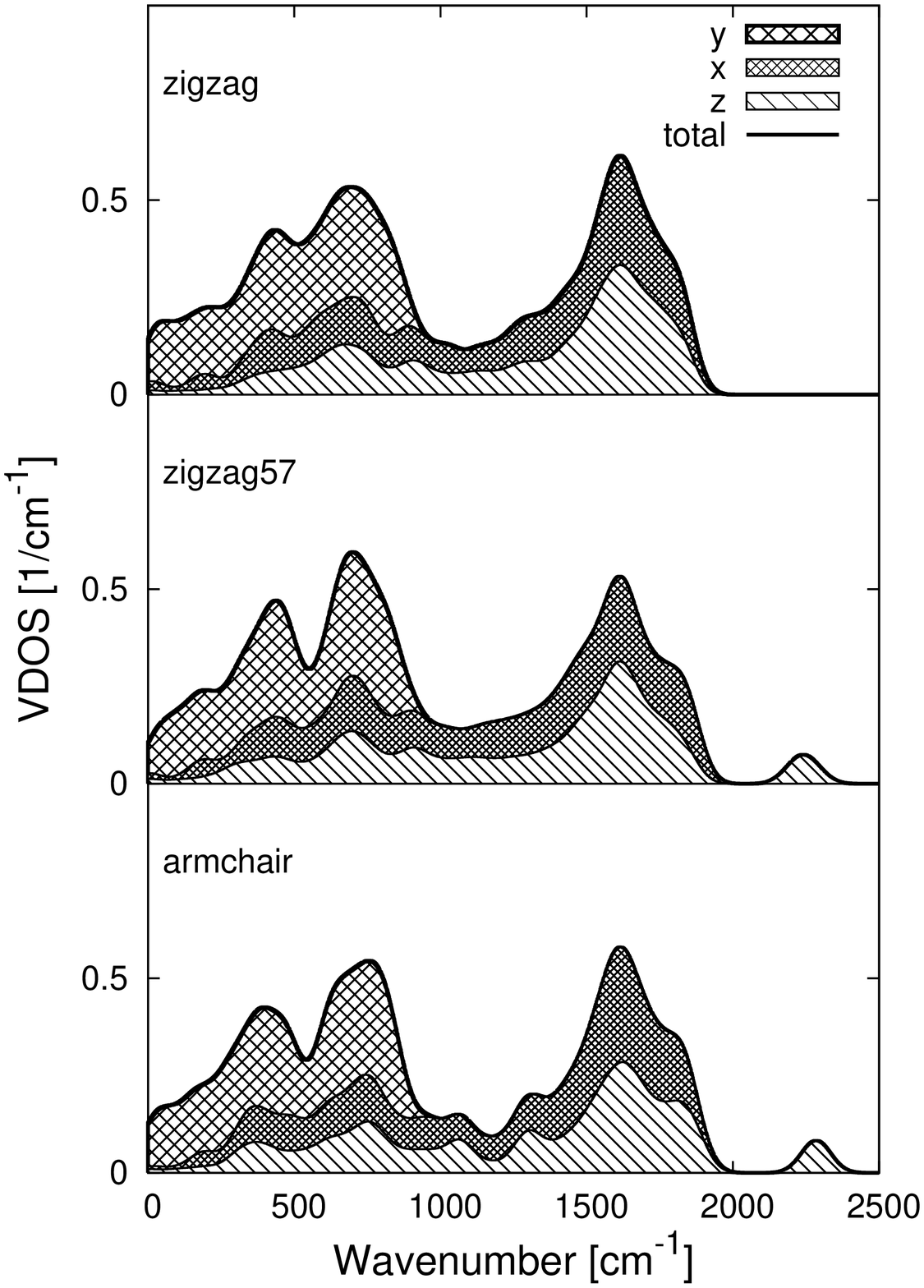}}
\caption{Total vibrational density of states of zigzag, reconstructed zigzag (zigzag57) and armchair graphene nanoribbons. Contribution of different directions in vibrational states are shown in different patterned areas. Vibrational density of states is broadened with normal distribution with $\sigma=45$~cm$^{-1}$, chosen to illuminate general changes in VDOS under reconstruction.}
\label{fig3}
\resizebox{0.55\vsize}{!}{\includegraphics{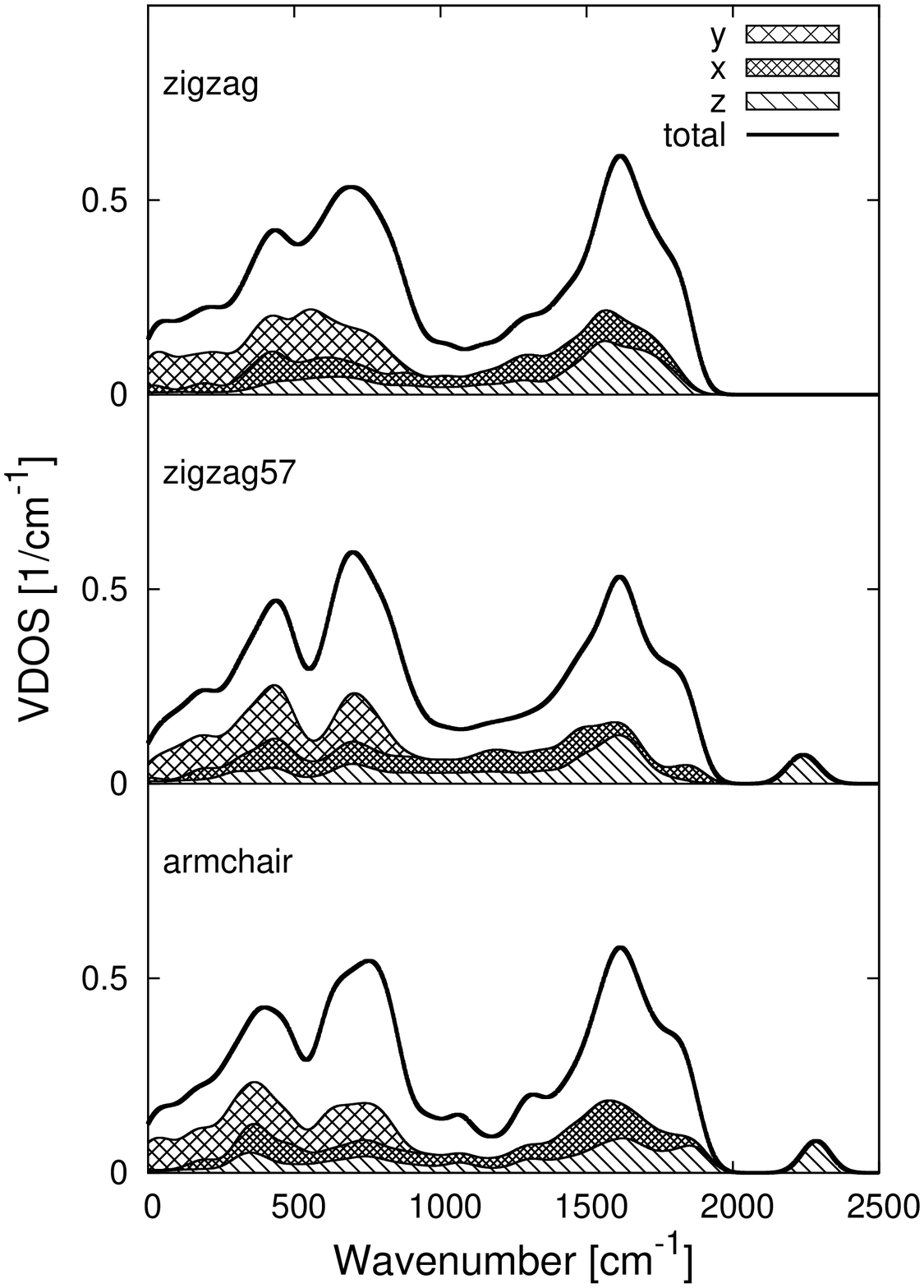}}
\caption{As figure \ref{fig3}, but showing the edge weighted vibrational density of states of zigzag, reconstructed zigzag (zigzag57) and armchair graphene nanoribbons.}
\label{fig4}
\end{figure}

We use density functional based tight-binding method to calculate forces\cite{frauenheim_JPCM_02}, to optimize structure\cite{bitzek_PRL_06} and to calculate the vibrational eigenmodes for the systems. For Raman spectra calculations we use non-resonant bond polarization theory\cite{bp1,bp2}, that has been used succesfully for ribbons\cite{APL_91_173108} and for non-identical carbon nanotubes\cite{vacancies_PRB_08,torsion,bend}. The slight consistent energy shift in the high-energy modes\cite{vacancies_PRB_08} does not affect the qualitative changes in Raman spectra or VDOS we want point out in this paper.

Figure \ref{fig1} shows the atom structures of zigzag, reconstructed zigzag (zigzag57) and armchair graphene nanoribbons. Widths and numbers of atoms of the ribbons are chosen to be easily comparable; also other sizes were systematically studied and all qualitative results given below were supported by these systematics. Edge atoms, as we define them, are drawn in gray and are used to determine edge-weighted vibrational density of states, where the edge-weight is ratio
\begin{equation}
w_{edge}^{\mu}=\frac{\sum_{i \in edge}{\bf v}_i^{\mu}\cdot {\bf v}_i^{\mu}}{\sum_{j \in all}{\bf v}_j^{\mu}\cdot {\bf v}_j^{\mu}} 
\end{equation}
where (3-dimensional vector) ${\bf v}_i^{\mu}$ is the component on atom $i$ for eigenmode $\mu$. Hence $w_{edge}^{\mu}=1$ means complete localization to edge, $w_{edge}^{\mu}=\frac{80}{200}$ means "uniformly distributed" mode, and $w_{edge}^{\mu}=0$ would mean a mode localized in the central region. Contributions of different directions are also analysed, analogously including weight of the components of the vibrational eigenvectors ${\bf v}_i^{\mu}$ in either $x$-, $y$- or $z$-direction,
\begin{equation}
w_x^{\mu}=\frac{\sum_{i }{\bf v}_{i,x}^{\mu}\cdot {\bf v}_{i,x}^{\mu}}{\sum_{j \in all}{\bf v}_j^{\mu}\cdot {\bf v}_j^{\mu}} 
\end{equation}
and similarly for y and z. Hence $w_x^{\mu}+w_y^{\mu}+w_z^{\mu}=1$ and $w_{x,edge}^{\mu}+w_{y,edge}^{\mu}+w_{z,edge}^{\mu}=w_{edge}^{\mu}$; $w_x^{\mu}$ measures how much of the mode is in the x-direction and $w_{x,edge}^{\mu}$ measures how much of the mode at the edge is in the x-direction.

Figure \ref{fig2} shows the xx,xz and zz polarized Raman spectra of the ribbons, where e.g. xz stands for incident light polarization in x- and scattered light polarization in z-direction. The polarization pictures xx and zz show the same modes, but with different intensities. Polarization pictures that contain y-direction and out-of-plane modes yield tiny intensities and have been left out from figure \ref{fig2}. Orientation of the ribbons with respect to coordinate axis is shown in figure \ref{fig1}. The most pronounced modes in fi\-gure \ref{fig2} include breathing modes (intensive low-energy peaks in xx/zz spectra), G-band-related high-energy modes (intensive high-energy peaks in the xz and xx/zz spectra) and edge-localized modes (symbols (L) in xz and zz spectra). One of the main points is that only edge-localized modes undergo observable changes under reconstruction, others being within $2$~cm$^{-1}$ from corresponding zigzag ribbon modes. These changes in edge-localized Raman active modes make the reconstruction visible. For zigzag57 the edge-localized mode of zigzag edge at $1550$~cm$^{-1}$ disappears in xz polarization spectrum in figure \ref{fig2} and edge-localized triple-bond vibration becomes visible at energies around $2250$~cm$^{-1}$ in zz polarization spectrum of zigzag57 ribbon, like in armchair ribbon. Intensity of the triple-bond vibration in reconstructed zigzag ribbon is smaller in our bond polarization theory than in armchair ribbon, mainly because of the differences in bond angles on the edge (the edge profile of zigzag57 is more linear).

Couple of new Raman active modes can be noticed to appear for reconstructed zigzag edge, assinged with symbols (1) in xx and (2) in xz polarization spectrum in fi\-gure \ref{fig2}. Mode (1) at $1500$~cm$^{-1}$ includes stretching of edge bonds of pentagons. Mode (2) at $1600$~cm$^{-1}$ is mainly a longitudinal mode where first normal zigzag row of atoms in the zigzag57 ribbon vibrates as the edge-localized mode of zigzag ribbon. Neither of the modes (1) nor (2) is fully localized on the edge.
 
Breathing modes remain unchanged, because they are collective vibrations which hold the edges internally unchanged during the vibration. Similarly, because G-band-related high-energy modes vanish towards the edge, reconstruction has only minor effect on them. Raman active out-of-plane vibrations have much lower intensity than in-plane vibrations and are not therefore considered here.

Next we compare more generally the vibrational pro\-perties of the different ribbon edges. Figure \ref{fig3} shows the total vibrational density of states, analysed with contributions from different directions. The surprising and signi\-fi\-cant result is that zigzag57 ribbon's vibrational density of states resembles armchair ribbon's rather than zigzag ribbon's vibrational density of states. This means that for vibrational modes it is not the orientation of the underlying honeycomb lattice that is important, but only the \emph{local structures of the edges}. Comparison of changes bet\-ween total vibrational density of states in figure \ref{fig3} and edge-weighted vibrational density of states in figure \ref{fig4} reveals that the most significant changes are seen in edge-localized modes. Triple-bond vibrations in z-direction on the high energies appear as a separate peak around $2250$ cm$^{-1}$ in figures \ref{fig3} and \ref{fig4} for both reconstructed zigzag ribbon and armchair ribbon. The triple-bond edge modes in z-direction in the zigzag57 ribbon, somewhat deplete modes from the energy range $1700-2000$~cm$^{-1}$ of the G-band-related modes in z-direction. At the same time in the same energy range the number of edge-related in-plane vibrations in x-direction are increased due to reconstruction - induced edge stiffening. Similar effects are seen even in lower energies. Some edge-localized out-of-plane modes around $500-650$~cm$^{-1}$ are upshifted in energy for zigzag57 ribbon for the same reason. Even the energy of the lowest energy out-of-plane modes increase, which can dramatically alter low-temperature properties (such as heat capacity) of the ribbons.

Hence, due to edge stiffening, edge vibrations gene\-rally up-shift in energy under reconstruction. This type of changes in mechanical properties can have an effect on applications such as sensors. The similarity of armchair and zigzag57 ribbon's VDOS is general observation, as the resemblance and also other qualitative features of the vibrational properties remained also for wider ribbons (up to $W=40$~\AA).

To conclude, we have analysed the Raman spectra and vibrational density of states of the recently reported self-passivating edge reconstruction of zigzag ribbons. In the Raman spectra edge-localized triple-bond vibrations become visible like the corresponding mode in armchair ribbon's spectra. These high-energy vibrations can be used to identify the reconstruction by concentrating the Raman measurement on the edge.

The predicted intensity differences between triple-bond vibrations of the zigzag57 ribbon and the armchair ribbon could be used to separate the reconstructed edge structure from the armchair edge structure with help of scanning tunneling microscopy images. Scanning tunneling microscopy can also help in avoiding identification difficulties with ribbons that have combination of zigzag and armchair edges. For those ribbons the density of triple-bonds at the edges is lower and further intensity of the triple-bond vibrations is lower like in reconstructed zigzag ribbon, which can lead to mix up between these structures.

The rest of the visible Raman active modes of the ribbons do not change for zigzag57 and the breathing mode has the same ribbon width dependence. Of course, in very narrow ribbons with majority of atoms on the edge, ener\-gies or spatial nature of the G-band-related high-energy modes may still change.

\begin{acknowledgement}
This research is supported by the Academy of Finland through the
FINNANO consortium MEP (molecular electronics and nanoscale photonics) and project 121701. S. Malola acknow\-led\-ges a grant from the Finnish
Cultural Foundation.
\end{acknowledgement}

\end{document}